# Retrieving the Red Edge on Earth-like Planets with Heterogeneous Clouds and Surfaces

Zachary Burr[1,2], Mario Damiano[1], Vincent Kofman[3], Renyu Hu[4,5,6,1], and Geronimo L. Villanueva[7]
[1] Jet Propulsion Laboratory, California Institute of Technology, Pasadena, CA 91109, USA; zaburr@phys.ethz.ch
[2] ETH Zurich, Institute for Particle Physics & Astrophysics, Wolfgang-Pauli-Str. 27, 8093 Zurich, Switzerland
[3] Centre for Planetary Habitability, Department of Geoscience, University of Oslo, 0371 Oslo, Norway
[4] Department of Astronomy & Astrophysics, The Pennsylvania State University, University Park, PA 16802, USA
[5] Center for Exoplanets and Habitable Worlds, The Pennsylvania State University, University Park, PA 16802, USA
[6] Institute for Computational and Data Science, The Pennsylvania State University, University Park, PA 16802, USA
[7] NASA Goddard Space Flight Center, 8800 Greenbelt Road, Greenbelt, MD 20771, USA

## Abstract

The detection and characterization of potentially habitable exoplanets is one of the chief goals of astrophysics for the coming decades. Imaging in reflected light is well suited for characterizing Earth-like planets, because much can be learned about these planets in this wavelength range (i.e., ∼0.3–2 $\mu$m). Several studies have been conducted to determine the abilities and limitations of reflectance spectroscopy, but most previous studies assumed a homogeneous atmospheric and surface composition. Here we investigate how heterogeneities in the atmosphere and surface of an Earth-like planet impact retrieval results. We extend the ExoReL$^{\Re}$ retrieval framework to include a step function for retrieving wavelength-varying surface albedo. We then use it to retrieve on visible to near-infrared spectra of realistic 3D Earth models with different surface features in view and varying cloud types/distributions synthesized with the Planetary Spectrum Generator. Including the ability to fit for wavelength-dependent albedo mitigates degeneracies that arise when using 1D models to analyze 3D planets, and we recover an Earth-like planet in all cases. We detect surface albedo steps at ∼0.7 and ∼1.1 $\mu$m despite clouds, both when significant lands are in view and when the spectra are averaged to account for a longer integration time. Our findings support the application of the vegetation red edge as a biosignature in the context of the Habitable Worlds Observatory. This study highlights the importance of considering a range of—particularly wavelength-dependent—surface albedos when using reflectance spectroscopy to characterize Earth-like exoplanets.

## 1. Introduction

The quest to discover and characterize potentially habitable exoplanets remains a central focus of contemporary space research. As outlined in the Astro2020 decadal survey, developing a ⩾6 m telescope capable of high-contrast imaging of habitable-zone terrestrial exoplanets in reflected light is a top priority (National Academies of Sciences, Engineering, and Medicine 2021). Such a mission would enable unprecedented insights into the abundance and composition of Earth-like planets in our interstellar neighborhood, providing crucial data on their atmospheres and potential habitability.

Previous research has explored Earth as an exoplanet analog. There have been several full-disk-integrated observations made of the Earth as if it were an exoplanet: using light reflected by the Moon (e.g., earthshine: L. Arnold et al. 2002; M. C. Turnbull et al. 2006) or using instruments on other spacecraft performing Earth flybys (e.g., EPOXI: N. B. Cowan et al. 2009). These observations demonstrate how Earth's reflected light spectrum fluctuates as Earth rotates.

Retrievals on both simulated and observed Earth spectra have shown the ability of reflectance spectroscopy to characterize Earth-like planets. Y. K. Feng et al. (2018) demonstrated the use of nested sampling retrievals on reflectance spectra to obtain constraints on key planetary parameters and explored the degeneracy between the planetary radius and surface albedo. T. D. Robinson & A. Salvador (2023) validated reflected light retrievals on real Earth data captured from EPOXI. M. Damiano & R. Hu (2022) showed the need for the near-infrared (NIR) portion of the spectrum to constrain the cloud properties and $CO_2$ and $CH_4$ gas abundances, for reflectance spectroscopy of Earth-like planets. A similar need for the UV spectrum to constrain $O_3$ abundances was demonstrated by M. Damiano et al. (2023). More recent studies have also investigated how a lack of prior information about the planet's mass (M. Damiano et al. 2025) and orbital parameters (A. Salvador et al. 2024) can cause degeneracies in the retrieval results. Grid-based retrievals are a complementary technique that also allow for determining atmospheric abundances from reflected light spectra but require fewer computational resources than nested sampling (N. Susemiehl et al. 2023). Grid-based retrievals have been used to determine the optimal spectral bands needed for identifying key biosignatures (N. Latouf et al. 2023, 2024).

In addition to atmospheric biosignatures, reflected light spectroscopy opens up the possibility of detecting surface biosignatures: spectral features caused by the presence of biology on the surface of the planet (J. Gomez Barrientos et al. 2023;

N. Parenteau et al. 2026). One such widely discussed surface biosignature is the vegetation red edge. The red edge is a sharp increase in reflectance at about 0.7 μm found in Earth vegetation due to the pigmentation of chlorophyll. The increase in surface albedo due to Earth's vegetation can be detected in Earth's reflectance spectrum (C. Sagan et al. 1993). It is reasonable to hypothesize that if other planets have photosynthetic organisms, such a red edge may be detectable there too.

It has been shown before that detecting the red edge from an Earth-like planet's reflectance spectrum could be challenging due to heterogeneous surfaces and varying cloud coverage (S. Seager et al. 2005). So far, the majority of studies of reflected light retrievals have only considered a model with a surface albedo that is homogeneous and constant with wavelength. The modern Earth has a relatively complex surface that is highly heterogeneous and has a wavelength-dependent albedo. The effect of this heterogeneity on reflected light retrievals is understudied. Recent studies have shown that a flat albedo retrieval model cannot accurately reproduce the red edge of Earth's vegetation, causing biases in the retrieved atmospheric abundances (F. Wang et al. 2022). J. Gomez Barrientos et al. (2023) took this a step further by demonstrating that these biases could be corrected and the spectral location of the red edge could be retrieved by using a parameterized albedo in the retrieval model. Models utilizing a linear combination of specific surface types have also been used to fit the wavelength-dependent surface albedo and to determine the fraction of land to ocean (A. G. Ulses et al. 2025). However, these previous works used an averaged albedo over the full surface, and did not consider the longitudinal and latitudinal distribution of different surface types or how this distribution changes as the Earth rotates. J.-N. Mettler et al. (2020, 2023, 2024) have demonstrated how Earth's emission spectrum varies both seasonally and spatially and how this variation affects retrievals, but to our knowledge, no similar study has been carried out for reflectance spectra.

It is therefore necessary to investigate how the heterogeneous surface and other time-of-day effects impact retrievals of an Earth-like planet's reflectance spectrum, and in particular, whether the red edge can still be detected when considering a more realistic distribution of land, water, and clouds over the surface. We adapt the ExoReL$^{\Re}$ retrieval framework (M. Damiano & R. Hu 2022; M. Damiano et al. 2023, 2025) to include a step function for fitting the wavelength-dependent albedo. We then apply this framework to spectra generated by the Planetary Spectrum Generator (PSG; G. L. Villanueva et al. 2018, 2022) of Earth as an exoplanet at various times of day (V. Kofman et al. 2024). We aim to contribute to the ongoing research of reflectance spectra of Earth-like planets by investigating how varying time of day, surface features, and cloud cover impact the retrieval of atmospheric properties and surface biosignatures.

## 2. Methods

For this study, spectra of a modern Earth-like exoplanet were simulated using PSG and then retrieved using ExoReL$^{\Re}$.

### 2.1. Input Spectra

The spectra analyzed are based on recent simulations of Earth as an exoplanet described by V. Kofman et al. (2024), where ground cover and atmospheric composition were ingested from MODIS and MERRA-2 climatology (R. Gelaro et al. 2017; M. Friedl & D. Sulla-Menashe 2022). The products of the moderate-resolution imaging spectroradiometer (MODIS) describe high-resolution ground coverage, which has been adopted into five reflectivity curves from the USGS (ocean, ice/snow, forest, grass, and soil; R. F. Kokaly et al. 2017). Sea ice and snow coverage were layered on top of the MODIS maps using information from the National Snow and Ice Center (W. Meier et al. 2021). MERRA includes high-spatial-resolution atmospheric data at 3 hr intervals. In this work, the spatially varying vertical profiles of $O_3$, $H_2O$, and liquid/ice clouds were adopted from the MERRA-2 database and considered in the simulations. The simulations also include the effect of ocean glint. While the spectra presented by V. Kofman et al. (2024) include additional sea salt aerosols, the ones used for this study only include liquid water and water ice aerosols (clouds). It was demonstrated by V. Kofman et al. (2024) that the methods reproduce the reflectivity of the Earth as seen from space on both a spatial and a spectral level. Each of the simulations consists of $2 \times 6500$ unique radiative transfer calculations, one set at a maximum cloud cover and one cloud-free. The individual spectra are combined using the cloud coverage fraction, and weighted appropriately to obtain the disk-averaged spectra shown below. For more details, see V. Kofman et al. (2024) and G. L. Villanueva et al. (2018, 2022).

V. Kofman et al. (2024) generated 18 spectra at nine different times of day with and without clouds. Only the cloudy spectra are used in this paper. The spectra are shown in Figure 1. Due to the rotation of the Earth, these spectra show an Earth with varying surface features, as well as differences in cloud cover. This allows us to understand how the heterogeneous surface affects our ability to reliably characterize Earth-like planets when considering their dynamic nature. All spectra were generated with the planet at quadrature. For this work, the wavelength range considered is from 0.3 to 2.0 μm, with a constant spectral resolving power of $R = 70$. This range and spectral resolution are in the range of values considered by previous mission concept studies (The LUVOIR Team 2019; B. S. Gaudi et al. 2020), as well as for the future Habitable Worlds Observatory.

Gaussian noise was added to the spectra using the noise model described by M. Damiano et al. (2025), with an assumed signal-to-noise ratio (SNR) of 20 at 0.75 μm. Additionally, the average of the nine cloudy spectra was taken to represent a spectrum obtained from an integration time longer than the rotation period of the planet. As the nine observations are not evenly spaced in time this is not an exact calculation of the Earth spectrum integrated over a full day; however, it still serves as a useful test case for an arbitrary Earth-like planet observed with a future direct imaging mission.

### 2.2. Retrieval Model

ExoReL$^{\Re}$ is a Bayesian inverse retrieval framework for spectra of exoplanets imaged in reflected light (M. Damiano & R. Hu 2020, 2022). A forward model (R. Hu 2019) generates spectra that are compared with the observations. Free parameters are systematically varied to determine the combinations that produce spectra statistically close to the observations (using the logarithm of a normal distribution as the log-

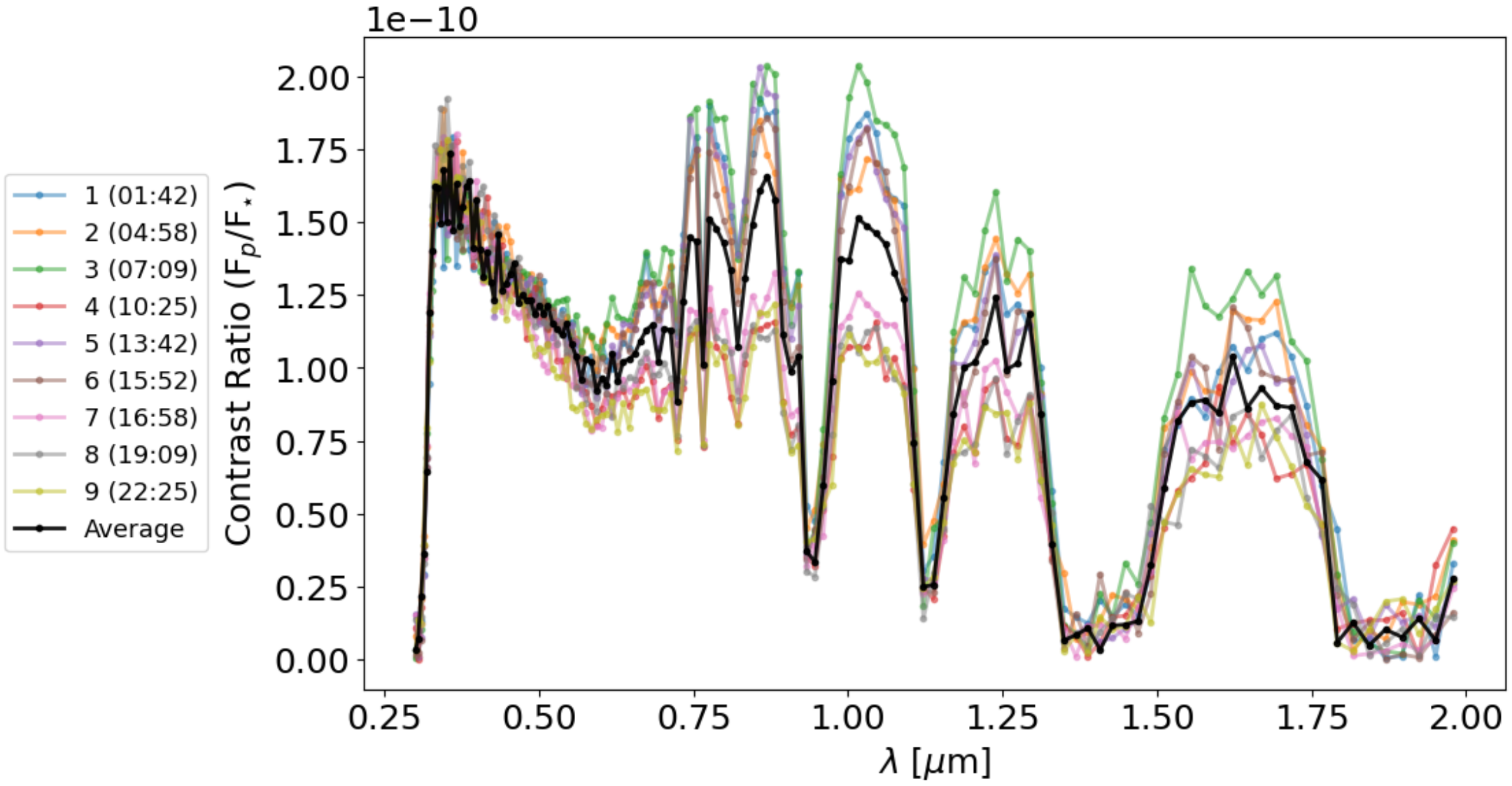

**Figure 1.** Input spectra generated with PSG from simulations of modern Earth. The spectra were generated at nine times of day, showing different surface features as the planet rotates (V. Kofman et al. 2024). A wide diversity in the spectra can be seen, demonstrating the wavelength dependence of the surface features. The black line is an average of all the spectra, representing an observation with a longer integration time. The data for all of the spectra used in this work are available in machine-readable format as the data behind the figure.

(The data used to create this figure are available in the online article.)

likelihood function). This allows for constraining the value of certain key parameters of the planet.

In this study, the partial pressures of $H_2O$, $CO_2$, $O_2$, $O_3$, $CH_4$, and $N_2$ are free variables, as described by M. Damiano et al. (2025). This allows all the gases to be fit independently, with no gas being preferred as the background gas a priori. This method for fitting the gas abundances for reflected light retrievals has already been demonstrated to give good results for Earth-like planets (S. Hall et al. 2023; A. Salvador et al. 2024). The surface pressure, $P_0$, is not fit directly but is derived from the sum of the partial pressures of all the gases. In the posterior distribution plots (e.g., Figure 5), the partial pressures of the gases are converted to the volume mixing ratio (VMR) for simplicity. Appendix B shows a comparison between retrievals using the partial pressure and the center log ratio of VMR as fit parameters.

We consider liquid water clouds as the only condensates. The clouds are modeled as described by M. Damiano & R. Hu (2020) and are fit with the cloud top pressure ($P_{\mathrm{top,H_2O}}$), cloud depth ($D_{\mathrm{cld,H_2O}}$), and condensation ratio ($CR_{\mathrm{H_2O}}$), as well as the VMR of water below the clouds. There is also a parameter for the cloud fraction, which sets what portion of the surface is covered by clouds. A two-column model is used, where one column is completely cloudy and one is cloud-free. The resulting spectrum is the average of these two, weighted by the cloud fraction. (For more details, see M. Damiano et al. 2025.) The (vertically constant) mean particle size for the cloud particles was likewise made a free parameter in ExoReL$^{\Re}$, as this was shown to give better results when retrieving on the more detailed cloud model used by PSG. The planet radius ($R_p$) is included as a free parameter, but the planet mass is fixed to 1 $M_{\oplus}$.

For the gas absorption lines we use the HITRAN2012 database (L. S. Rothman et al. 2013). The optical properties of the clouds are taken from G. M. Hale & M. R. Querry (1973).

**Table 1**
Model Parameters and Prior Probability Distributions Used in the Atmospheric Retrievals

| Parameter | Symbol | Prior |
|---|---|---|
| Partial pressure $H_2O$ (Pa) | PP($H_2O$) | $\mathcal{LU}(-7.0, 7.0)$ |
| Partial pressure $CH_4$ (Pa) | PP($CH_4$) | $\mathcal{LU}(-7.0, 7.0)$ |
| Partial pressure $CO_2$ (Pa) | PP($CO_2$) | $\mathcal{LU}(-7.0, 7.0)$ |
| Partial pressure $O_2$ (Pa) | PP($O_2$) | $\mathcal{LU}(-7.0, 7.0)$ |
| Partial pressure $O_3$ (Pa) | PP($O_3$) | $\mathcal{LU}(-7.0, 7.0)$ |
| Partial pressure $N_2$ (Pa) | PP($N_2$) | $\mathcal{LU}(-7.0, 7.0)$ |
| Cloud top pressure (Pa) | $P_{\mathrm{top,H_2O}}$ | $\mathcal{LU}(-7.0, 7.0)$ |
| Cloud depth (Pa) | $D_{\mathrm{cld,H_2O}}$ | $\mathcal{LU}(2.0, 7.0)$ |
| Condensation ratio | $CR_{\mathrm{H_2O}}$ | $\mathcal{LU}(-7.0, 0.0)$ |
| Cloud fraction | cld frac | $\mathcal{LU}(-3.0, 0.0)$ |
| Cloud particle size ($\mu$m) | $P_{\mathrm{size}}$ | $\mathcal{LU}(-1.0, 2.0)$ |
| Planetary radius ($R_{\oplus}$) | $R_p$ | $\mathcal{U}(0.5, 10.0)$ |
| Surface albedo 1 | $a_{\mathrm{surf,1}}$ | $\mathcal{U}(0.0, 1.0)$ |
| Surface albedo 2 | $a_{\mathrm{surf,2}}$ | $\mathcal{U}(0.0, 1.0)$ |
| Surface albedo 3 | $a_{\mathrm{surf,3}}$ | $\mathcal{U}(0.0, 1.0)$ |
| Wavelength albedo red edge [$\mu$m] | $\lambda_{\mathrm{surf}}$ | $\mathcal{U}(0.4, 0.8)$ |
| $\Delta$wavelength albedo NIR edge [$\mu$m] | $\Delta\lambda_{\mathrm{surf}}$ | $\mathcal{U}(0.01, 0.5)$ |

**Note.** $\mathcal{U}(a, b)$ is the uniform distribution between values $a$ and $b$, $\mathcal{LU}(a, b)$ is the log-uniform distribution between values $a$ and $b$.

The forward model is run at a spectral resolution of $R = 2000$, then binned down to match the input spectra. The model utilizes 180 atmospheric layers, which are equally spaced in log($P$).

The free parameters and their prior functions are listed in Table 1. ExoReL$^{\Re}$ uses `MultiNest` (F. Feroz et al. 2009) to sample the Bayesian evidence, estimate the parameters, and calculate the posterior distribution functions. `MultiNest` is used through its `Python` implementation `pymultinest` (J. Buchner et al. 2014). For all the retrieval analyses presented here, we used 1200 live points and 0.5 as the Bayesian evidence tolerance.

### 2.2.1. Fitting Wavelength-dependent Surface Albedos

It is well established that the Earth's heterogeneous surface can induce a strong wavelength dependence in the disk-averaged surface albedo (e.g., G. Roccetti et al. 2025), as different land types dominate the reflected signal toward the observer at different times. In particular, vegetation exhibits a distinctive spectral feature in the visible around 0.7 $\mu$m, where the albedo increases sharply across a narrow wavelength range, commonly referred to as the "red edge" (e.g., S. Seager et al. 2005). As the Earth rotates, the relative contribution of surface types such as desert, vegetation, and ocean changes, leading to time-dependent variations in the effective surface albedo.

In this study, we upgraded ExoReL$^{\Re}$ to allow for a nonflat, wavelength-dependent surface albedo, similar to the approach taken by J. Gomez Barrientos et al. (2023). Rather than adopting any prior knowledge of the Earth's specific surface properties, we chose a deliberately general and agnostic parameterization. The surface albedo is modeled as a step-like function, increasing the number of free parameters from one to five: three albedo values (before, across, and after the step) and two wavelength parameters that define the location and width of the transition in wavelength space (the first is an independent parameter that locates the position of a potential red edge, while the second is an additive parameter that, when combined with the first, determines the location of the step edge in the NIR). In the posterior distribution functions in Section 3, we show the sum of these two wavelength parameters to identify the NIR edge more easily. This flexible formulation is capable of capturing a sharp spectral transition consistent with a vegetation red edge, as well as a more gradual or nearly flat behavior representative of surfaces such as deserts or oceans, when warranted by the data.

### 2.3. 1D Test Run

In addition to the nine retrievals with varying surface features and cloud coverage, several "test run" retrievals were run on 1D spectra generated with PSG with and without clouds (i.e., with a homogeneous surface and atmosphere that do not vary with latitude and longitude, and a surface albedo that is constant with wavelength). The same noise model was used for these spectra; however, no Gaussian scatter was added to the spectra to ensure this does not bias the validation. Other than the surface, the inputs are identical to the other retrievals performed for this study. This was done as a check to show whether ExoReL$^{\Re}$ returns similar results on spectra generated with PSG without considering the heterogeneous surfaces and clouds. This could help in determining whether any retrieval discrepancies are due to the variations in surface and cloud cover or to other differences between the forward models. A surface albedo of 0.05 was chosen, which is roughly consistent with a 100% ocean planet. For these test runs, we also use a 1D retrieval model (i.e., with only one free parameter for the surface albedo, which is constant in wavelength).

The posteriors for these test runs and a more in-depth explanation of the results can be found in Appendix A. The validation shows that—in the absence of clouds—the ExoReL$^{\Re}$ framework accurately retrieves on spectra produced by the PSG model. Slight differences in the cloud models (in particular, the inclusion of ice cloud particles in the PSG model) can cause some remaining biases in the retrieval results.

## 3. Results

The full results for all nine retrievals are shown in animated form in Figure 2. The rest of this section will look at the aggregated results to determine the main impacts of the heterogeneous surface and other time-of-day effects on the retrieved parameters.

### 3.1. Retrieved Surface Albedo

Figure 3 displays the posterior distributions and correlations for $P_0$, the three surface albedos, and the planet radius for all retrievals. The retrieval results show clear sensitivity to the surface features, as the posteriors form distinct groups based on how much land is in view during that observation, and the retrievals also correctly identify that the albedo is higher when there is more land in view. When there is sufficient land in view, $a_{\mathrm{surf},2}$ is found to be greater than $a_{\mathrm{surf},1}$, demonstrating the influence of the red edge.

The retrieved wavelength-dependent albedos are shown more clearly in Figure 4. The retrievals fit the shape of the red edge, with an increase in albedo at $\sim$0.7 $\mu$m and a decrease around 1.1–1.2 $\mu$m. The effect is much more pronounced for the retrievals with more land in view, demonstrating a robust detection of the red edge, and the albedo is overall higher, demonstrating that ExoReL$^{\Re}$ is accurately retrieving the difference in albedo due to the differing surface features. The red edge cannot be confidently detected in the spectra with majority ocean. Although the best fit suggests a small increase in albedo due to the red edge, the posteriors show that it is also consistent with there being no red edge. However, all retrievals but one correctly place the transition wavelength at 0.7 $\mu$m, exactly where the red edge comes into effect, showing weak evidence for the red edge even for those retrievals with majority ocean.

The surface pressure is well constrained to be around 1 bar in all tested cases, i.e., the retrieval can accurately determine that the planet is rocky with a thin atmosphere. However, the majority of the retrievals are somewhat biased on radius, either with a smaller radius and a (nonphysically) high albedo or with a larger radius and a low albedo. This is because there is a strong correlation between the radius and the albedo. This correlation is quadratic: the reflectance increases with the square of the radius but increases linearly with albedo. Even with the wavelength-dependent albedo, this strong degeneracy means that it is difficult to correctly determine both the surface albedo and the radius at the same time.

### 3.2. Retrieved Atmospheric Abundances

The posteriors for the VMR of each gas for all retrievals are shown in Figure 5, color-coded by the amount of land in view. The posteriors are fairly consistent across all retrievals, with some variation due to the changing surface features. Eight out of nine of the retrievals correctly found an $N_2$-dominated atmosphere with a significant mixing ratio of $O_2$ (with $\log(\mathrm{VMR}(O_2)) \gg -3.5$, or approximately the Proterozoic $O_2$ abundance on Earth, T. W. Lyons et al. 2014). The posteriors for $O_3$ vary slightly, but the average over all retrievals is centered around the truth. For all retrievals, the surface pressure is between $10^4$ and $10^6$ Pa. With the simple

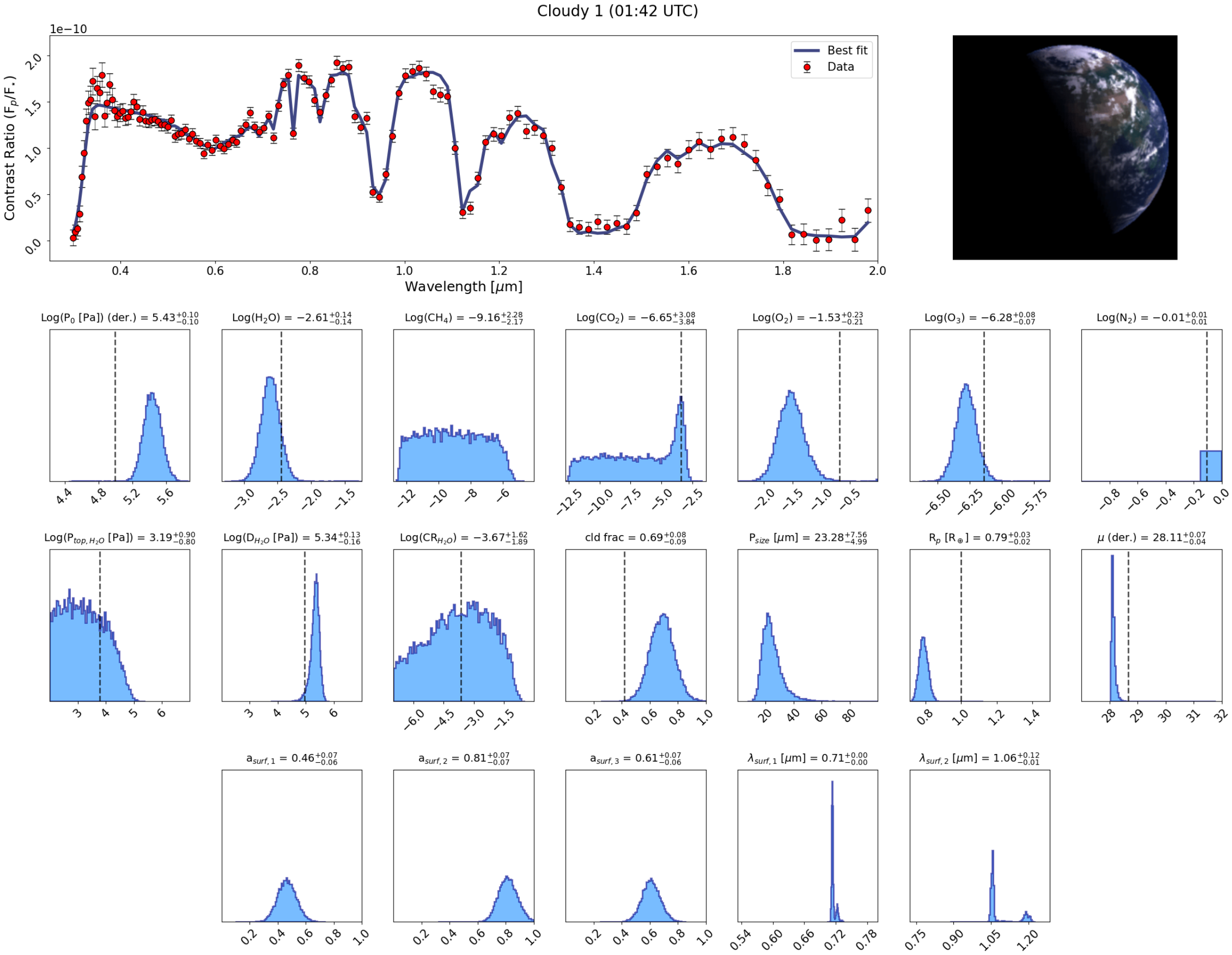

**Figure 2.** An animation showing the results of all nine retrievals. The top right shows the observer's view of the Earth at that time step. The dashed lines are the true values for these parameters. The $H_2O$ truth value is the VMR at the surface (which is what is reported by ExoReL$^{\Re}$). For $O_3$, which varies vertically in the PSG simulations but not in ExoReL$^{\Re}$, the true value is averaged over altitude. The spectrum and posteriors vary significantly between different time steps as different surface features rotate in and out of view. The real-time duration of the animation is 18 s.

(An animation of this figure is available in the online article.)

wavelength-dependent surface albedo formulation currently adopted by ExoReL$^{\Re}$ the surface type does seem to have a minor effect on the retrieved atmospheric abundances and surface pressure. This suggests that the wavelength-dependent albedos of the various surface types could be separated from the absorption by the atmospheric gases to a satisfactory extent.

The estimation of $O_2$ seems to be somewhat biased to either higher or lower abundances. At this spectral resolution ($R = 70$), there is only one data point within the main $O_2$ absorption feature at 0.76 $\mu$m, and the spectrum could be consistent with a wide range of $O_2$ abundances. The underestimation of $O_2$ is more prominent for retrievals with more land and also more vegetation in view. Due to the red edge, the albedo is much higher around the main $O_2$ absorption feature, raising that data point. Although the retrieval does a good job in retrieving the increase in surface albedo due to the red edge, even a slight difference between the retrieved and true albedos could obscure part of the $O_2$ feature, making it harder to fit. A similar occurrence happens with $H_2O$, as some $H_2O$ features are also obscured by the red edge: thus $H_2O$ is also slightly underestimated when there is more land in view.

The retrievals with more land in view also return some tentative evidence for $CO_2$, though with a very long tail such that the results are also consistent with no $CO_2$. Those retrievals with majority ocean in view would have no such hint of $CO_2$. $CO_2$ at the current abundance is known to be challenging to detect in this wavelength range, as the absorption features in the NIR are not particularly strong. The tentative evidence for $CO_2$ for the cases of majority land may be due to it being conflated with the decrease in surface albedo in the NIR on the other side of the red edge, which occurs near the $CO_2$ absorption features. Or it may be due to the overall increased flux making the absorption features more prominent. There was no $CH_4$ present in the input spectra, and the retrievals all place an upper bound on the $CH_4$ abundance of $10^{-5}$ regardless of the amount of land in view.

### *3.3. Biases Due to Not Fitting the Wavelength-dependent Albedo*

It is useful to also explore what biases could arise when not fitting for the wavelength-dependent albedo. Retrievals were also run on all nine input spectra with the same model but using only one free parameter for the albedo, which is constant at all wavelengths. This surface albedo parameter also has a uniform prior on [0.0, 1.0]. Additionally, these retrievals do not include the particle size as a free parameter, instead calculating it self-consistently as a function of pressure (as shown by M. Damiano et al. 2025). The results of these retrievals are shown in Figure 6.

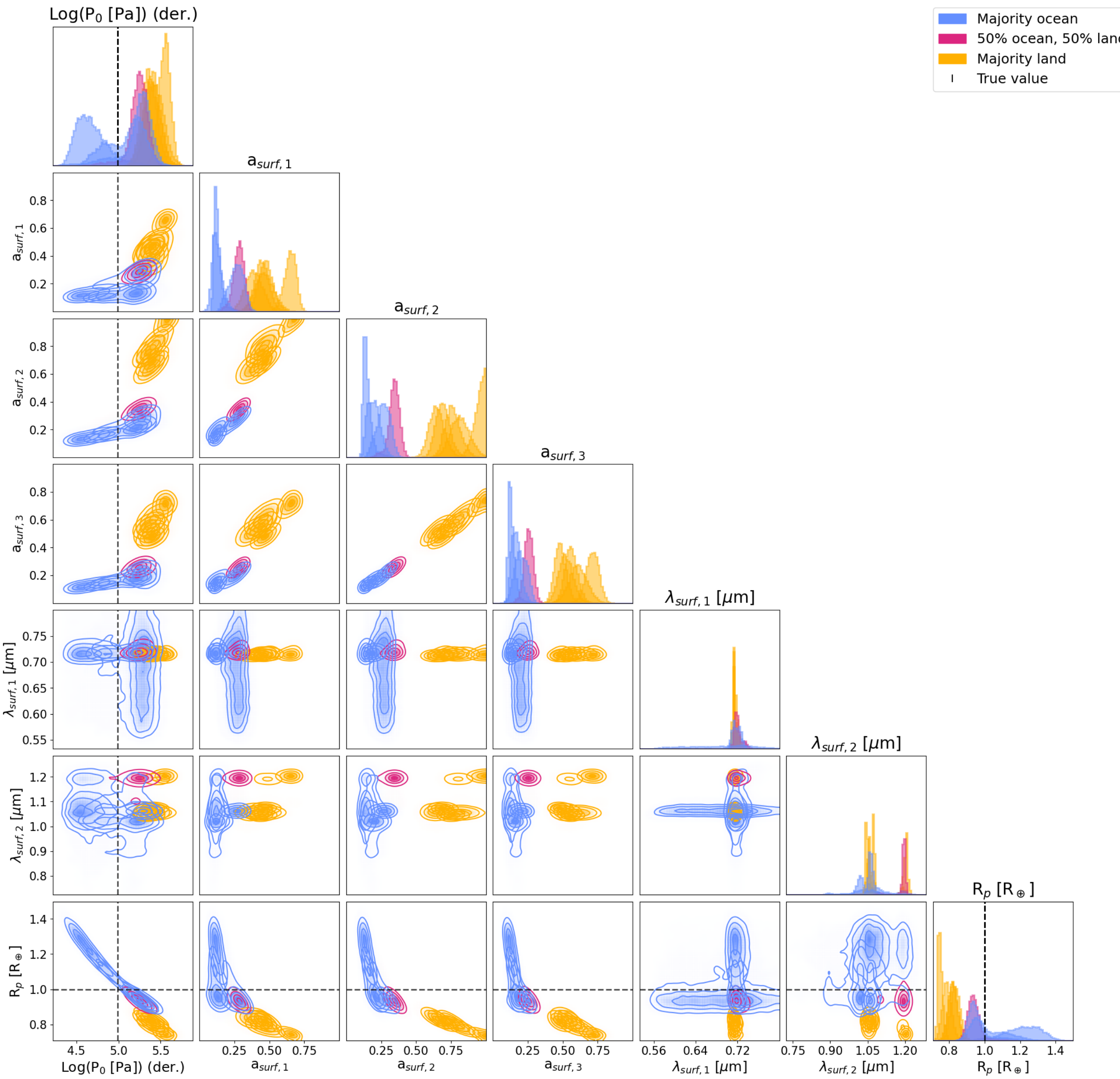

**Figure 3.** Posterior distributions for $P_0$, $a_{\rm surf}$, and $R_p$ for all retrievals, color-coded by the amount of land in view. Note that $P_0$ serves as a proxy for all the atmospheric gases as it is a combination of the partial pressures of each gas. There is often a strong correlation between radius and surface albedo, which makes it challenging to fit both together.

It is immediately clear that these retrievals perform far worse than the retrievals where the albedo step function is used. For $O_2$, the retrievals for majority land more severely underestimate the abundance, while all of the retrievals for majority ocean return an $O_2$-dominated atmosphere. One retrieval for majority land returns a potential detection of $CH_4$, and several others have nonzero constraints for $CH_4$, despite the fact there is no $CH_4$ in the input spectrum. This is likely due to the fact that red edge decreases again in the NIR, around the same region where there are $CH_4$ absorption features. Thus, the retrievals use $CH_4$ to try to fit this effect. Likewise, several retrievals for majority land return a confident detection of $CO_2$, which is likely spurious, as there should be little detectable $CO_2$ at this abundance and SNR, due to the $CO_2$ features being relatively weak in this range. This can be attributed to the same effect as the $CH_4$, as $CO_2$ also absorbs in a similar wavelength range. However, the false detection of $CO_2$ and $CH_4$ may also be partially attributed to the effect of the ice cloud particles, which have a similar impact of changing the slope of the spectrum in the 1.4–1.7 $\mu$m region (see Appendix A). The degeneracy between the radius and surface albedo seems to be more prominent here as well, with none of the retrievals returning the correct planetary radius.

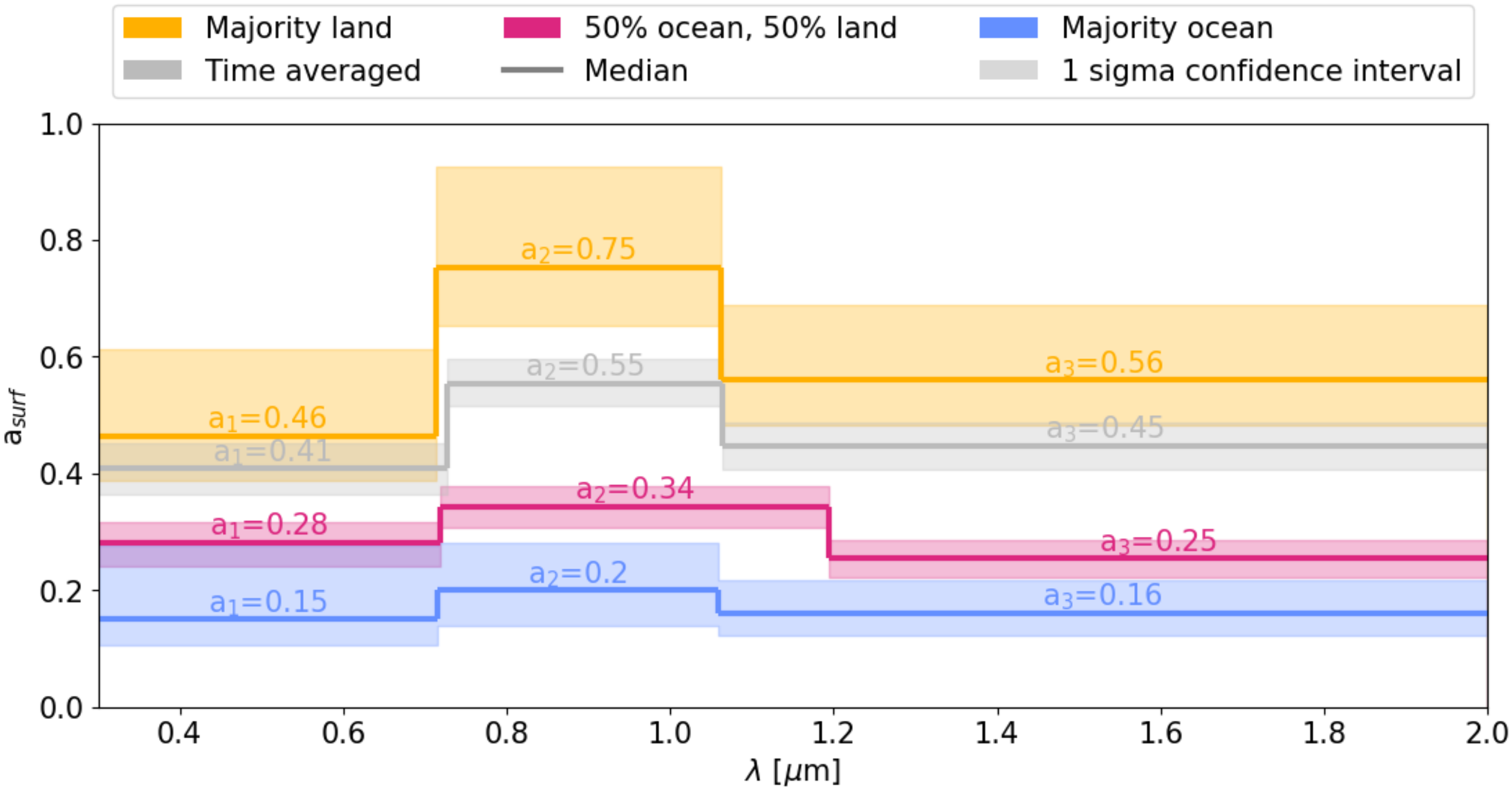

**Figure 4.** Retrieved median surface albedos versus wavelength for the combined posteriors of the three surface types. The retrieved surface albedo is considerably higher for those cases with more land in view, and the red edge signature is more prominent, which matches expectations. For the cases of majority ocean, the albedo is almost flat with wavelength, which is consistent with the fact that the reflectivity of water does not change much at these wavelengths. All of the retrievals place the red edge at the correct wavelength (∼0.7 μm), showing a robust detection.

This highlights the insufficiency of utilizing a flat albedo model to fit a planet with a (strongly) wavelength-dependent surface albedo. It also shows the tendency of retrievals to use other model parameters, such as absorption of $CH_4$ and $CO_2$, to attempt to fit effects that are not included in the model.

The preference for the wavelength-dependent albedo is also supported by the difference in the Bayesian evidence of the retrievals (Table 2). The models that include the wavelength-dependent albedo are strongly preferred over those with a constant albedo in nearly all cases. There are two retrievals where the wavelength-dependent albedo is only moderately preferred ($0 < \Delta \log Z < 5$), which are both cases with majority ocean where the effect of the red edge is minimized. This lends further support to a robust detection of the red edge in cases with majority land in view.

**Table 2**
Bayesian Evidence (logZ) for the Models with and without the Wavelength-dependent Surface Albedo for All Runs

| Input Spectrum | λ-dependent | Constant | ΔlogZ |
| --- | --- | --- | --- |
| 1 (Land) | 3179 | 3022 | 157 |
| 2 (Land) | 3181 | 3082 | 99 |
| 3 (Land) | 3167 | 2988 | 179 |
| 4 (Ocean) | 3204 | 3200 | 4 |
| 5 (Land) | 3147 | 2939 | 208 |
| 6 (Land) | 3159 | 3032 | 127 |
| 7 (50/50) | 3208 | 3173 | 35 |
| 8 (Ocean) | 3197 | 3194 | 3 |
| 9 (Ocean) | 3210 | 3185 | 25 |
| Time average | 3194 | 3103 | 91 |

**Note.** In almost all cases the model with wavelength-dependent albedo is very strongly preferred. For those cases with majority ocean in view, the model with wavelength-dependent albedo is only moderately preferred.

### *3.4. Time-averaged Spectrum*

The retrieval results from the time-averaged spectrum are shown in Figure 7. The posteriors are comparable to other retrieval results that are from the individual time steps—in particular, those retrievals with more land in view—and help support the same conclusions. The red edge is correctly retrieved, though less strongly than for those cases with majority land, because it has been moderately averaged out. The radius and $O_2$ abundance are somewhat underestimated, similar to those cases with majority land. Interestingly, this retrieval shows some hint of $CH_4$, despite there being no $CH_4$ in the input spectrum. This is likely due to the large decrease in albedo from the visible to the NIR: as the decrease happens close to the $CH_4$ features, it is possible that it is partly misidentified as $CH_4$. Nonetheless, the results are still consistent with no $CH_4$ being present, given the presence of the long tail in the 1D marginalized probability distribution function.

## 4. Discussion

### *4.1. On the Detection of the Red Edge*

Figure 4 already demonstrated our ability to infer the surface albedo from reflected light spectra. To further validate this capability, we compare retrievals that (i) allow for wavelength-dependent surface reflectivity using five independent albedo parameters and (ii) enforce a single, wavelength-independent ("gray") albedo parameter. We quantify the relative performance of these two models with a leave-one-out (LOO) cross-validation (CV) analysis following L. Welbanks et al. (2023). LOO compares models at the level of individual data points. For each spectral data point, the model is conditioned on all remaining points and then asked to predict the withheld point; this yields the expected log pointwise predictive density (elpd), a per-wavelength measure of out-of-sample predictive accuracy. Figure 8 reports the pointwise difference $\Delta \text{elpd} = \text{elpd}_{5A} - \text{elpd}_{1A}$, where positive values indicate that the model with five albedo parameters provides a better predictive description of that datum than the model with a single albedo. The largest positive differences occur in the 0.7–1.1 $\mu$m range, where the two albedo parameterizations diverge most strongly, indicating that this portion of the spectrum is not well captured by a single average albedo. Notably, several points outside the central albedo step also favor the five-parameter model, suggesting that the added flexibility leads to a redistribution of the inferred albedo across bands rather than merely improving the fit locally.

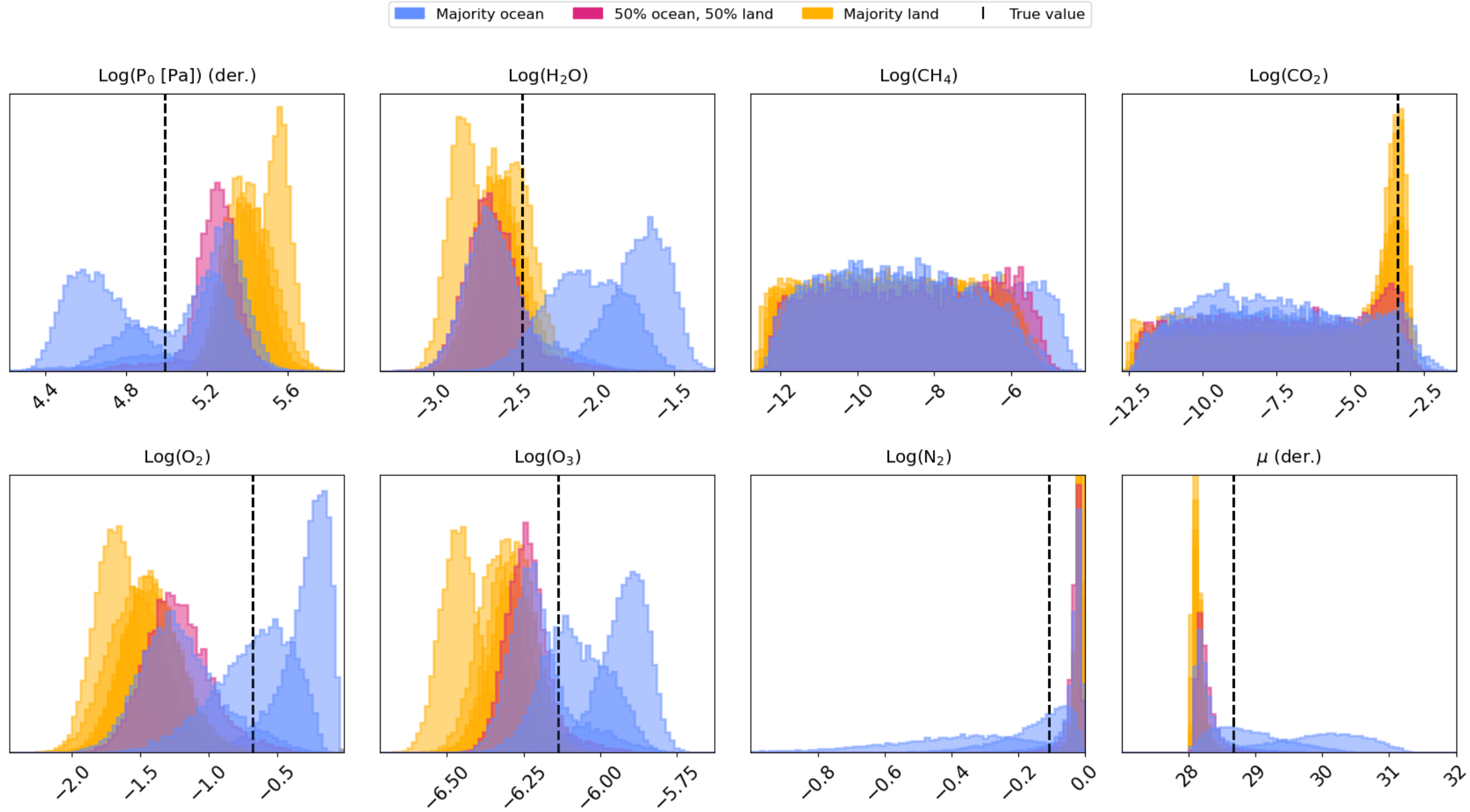

**Figure 5.** Posteriors for all atmospheric gases for all nine retrievals, color-coded by how much land is in view during that observation. The true value for $H_2O$ is the value at the surface, which is what is reported by the retrieval; the value of $O_3$ is the average of all atmospheric layers. The other gases do not vary vertically. The results show that there is a large difference between the posteriors depending on how much land is in view during the observation, with $O_2$ being underestimated when there is more land in view.

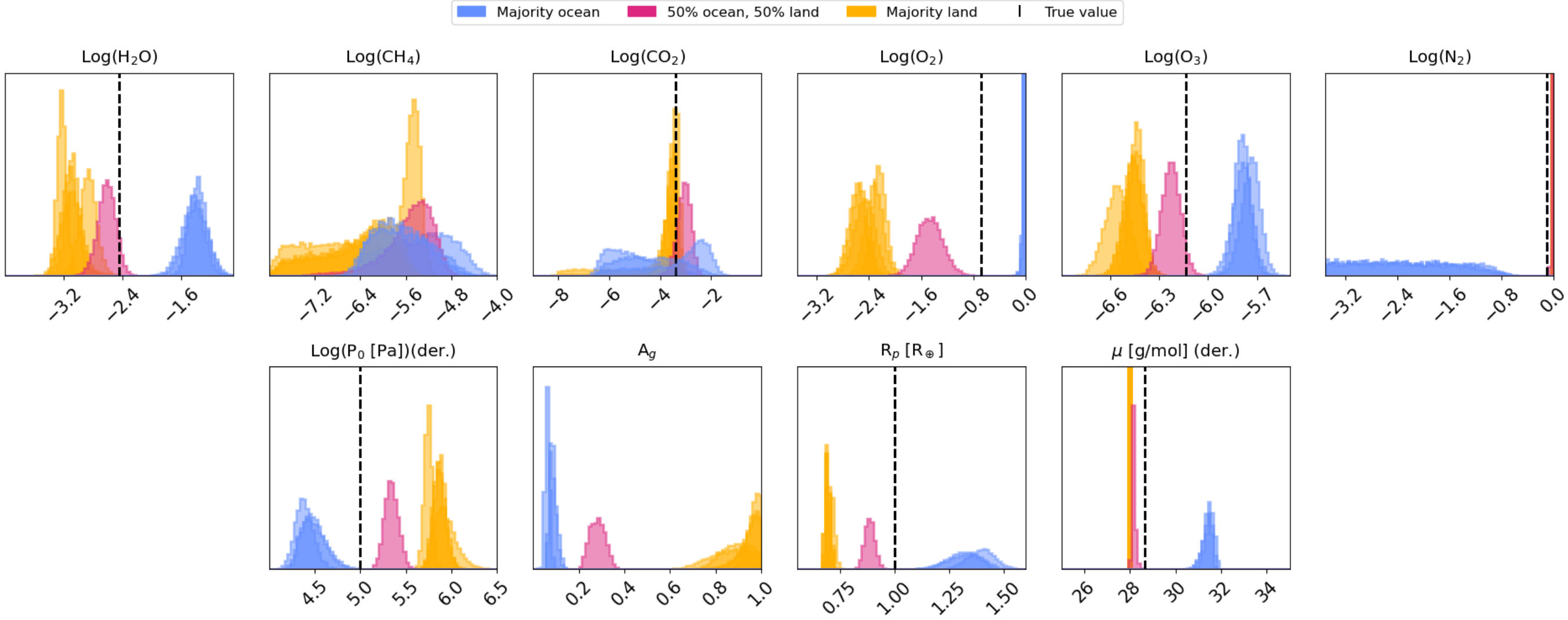

**Figure 6.** Posteriors for atmospheric gases, albedo, and radius for all retrievals when not fitting a step function for the albedo. The results are significantly worse than when the step function is utilized. The $O_2$ abundance is further biased in both directions. There are sometimes spurious detections of $CO_2$ and $CH_4$, and none of the retrievals return the correct radius.

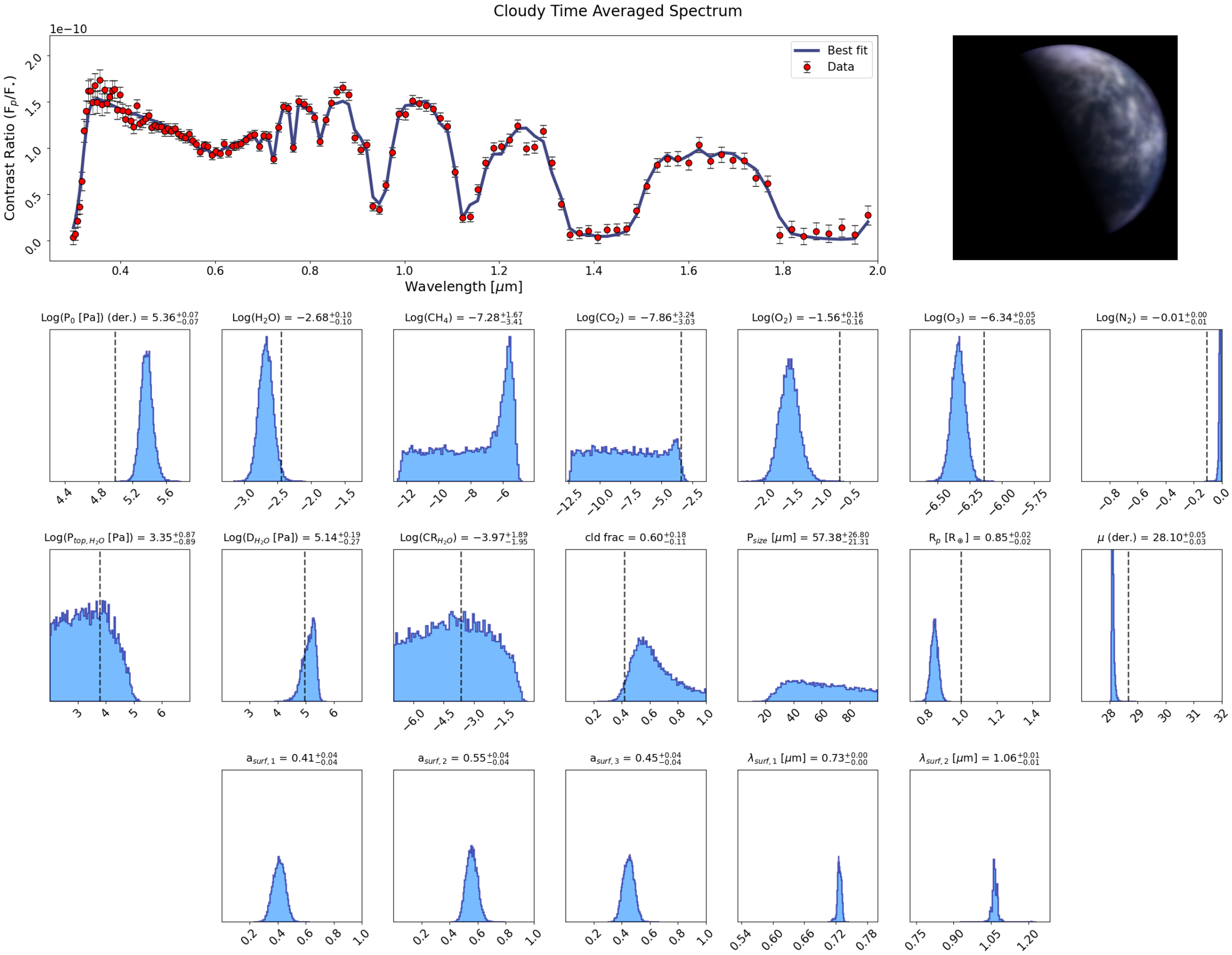

**Figure 7.** Best-fit spectrum and posteriors for the time-averaged spectrum. The results are in line with those of the individual retrievals: the retrieval identifying the presence of clouds (but not being able to accurately constrain them), determining $N_2$ is the dominant gas, and the red edge causing an underestimation in $H_2O$ and $O_2$ abundance. The constrained wavelength-dependent albedo is also shown in Figure 4 and lies between the values from the land- and ocean-dominated cases.

The results presented here demonstrate not only sensitivity to the surface albedo, but also the robust and repeatable identification of the vegetation red edge in scenarios where more than 50% of the surface is composed of land. This constitutes a key result, as it supports the detectability of vegetation as a potential biosignature in reflected light observations (S. Seager et al. 2005; J. T. O'Malley-James & L. Kaltenegger 2018, 2019; N. Parenteau et al. 2026). Crucially, these results show that the surface spectral signature can be disentangled from the wavelength-dependent effects of the atmosphere such as gas absorption, Rayleigh scattering, and the impacts of clouds. Being able to accurately separate the spectral signatures of the surface and atmosphere is a fundamental requirement for reliably interpreting surface habitability indicators and has important implications for the remote search for life beyond the solar system.

We utilize a similar approach to J. Gomez Barrientos et al. (2023), with a simple step function for the albedo. This approach does not assume any prior knowledge of the surface composition and is instead fully data-driven, allowing the surface properties to be inferred directly from the observed spectra. This makes the method broadly applicable to planets with unknown or non-Earth-like surfaces. By contrast, A. G. Ulses et al. (2025) adopt a physically motivated but more restrictive surface model, parameterizing the surface reflectance as a linear combination of empirically derived Earth surface components with fractional coefficients treated as free parameters. While this approach enables a direct mapping between retrieved parameters and known terrestrial surface types, it inherently relies on an Earth-centric basis set. The two methods are therefore complementary: empirical linear-combination models provide interpretability when Earth analogs are expected, whereas our formulation offers greater generality and flexibility, making it better suited for the characterization of truly unknown exoplanetary surfaces and for minimizing model-dependent biases when searching for biosignatures in diverse planetary environments.

### *4.2. What Happens if We Do Not Account for the Surface?*

This study shows the biases that can arise from using 1D models with a surface albedo that is constant with wavelength for retrieving on the reflected light spectra of terrestrial exoplanets. For several of the gases, the posteriors could be substantially biased, and the results of this study show that flat albedo models could struggle to retrieve key atmospheric and planetary parameters from spectra of an Earth analog with a heterogeneous surface.

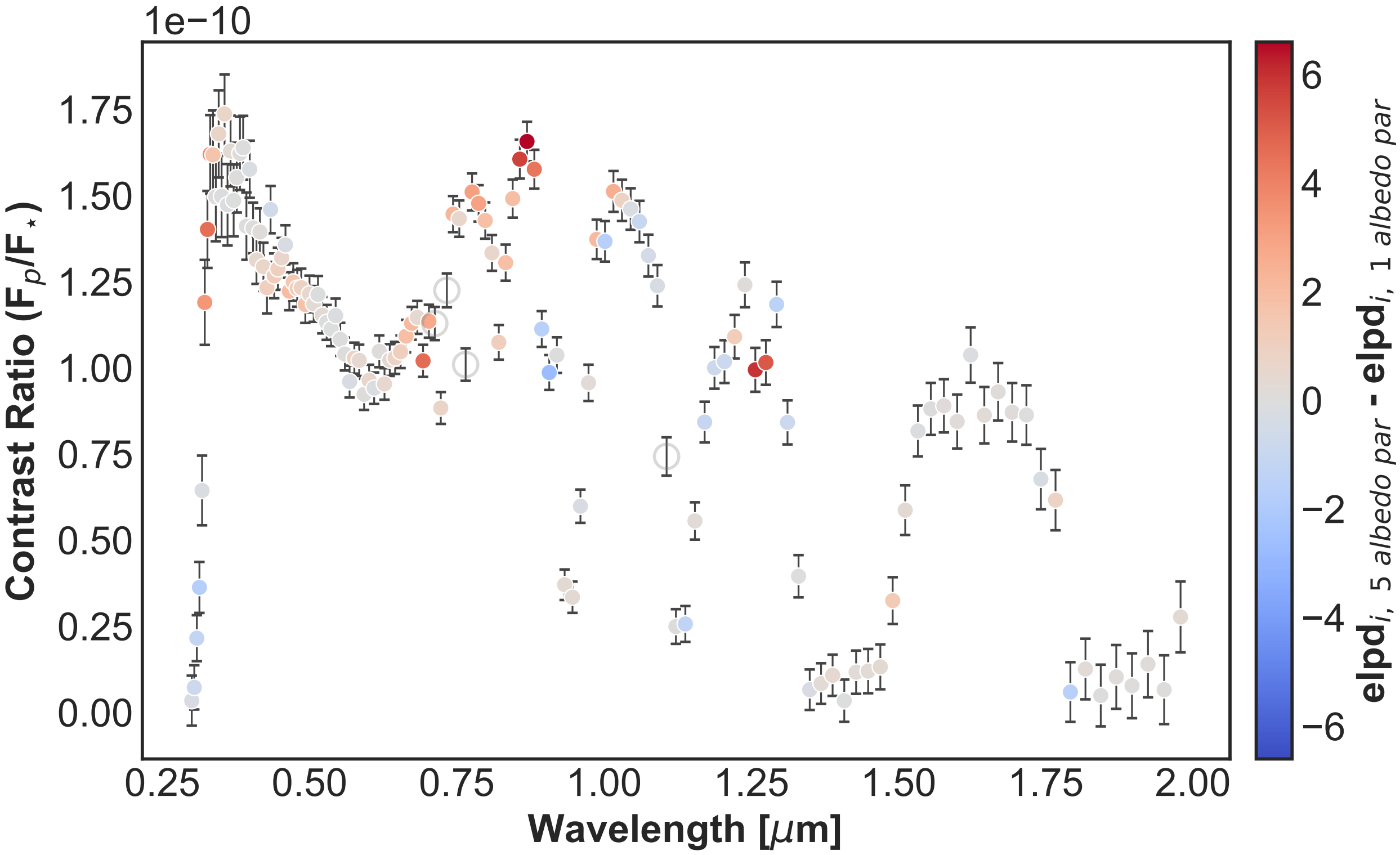

**Figure 8.** Pointwise leave-one-out cross-validation comparison between a wavelength-dependent surface albedo model (five albedo parameters) and a gray albedo model (single albedo parameter). Points show the measured planet-to-star contrast spectrum ($F_p/F_\star$) as a function of wavelength; the color encodes the per-point difference in expected log pointwise predictive density, $\Delta$elpd = elpd$_{5A}$ − elpd$_{1A}$. Positive (red) values indicate data points better predicted by the five-parameter albedo model, while negative (blue) values favor the single-parameter model. The data points with the gray empty circle had a Pareto $k$ value above 0.7; therefore the approximation and the elpd calculation could not be trusted and they are therefore excluded from the graph.

In particular, with a flat albedo model, the retrieval may try to use available free variables to fit some of the wavelength-dependent features from the surface. For example, the red edge due to vegetation could cause $O_2$ to be underestimated, or the cloud fraction to be retrieved as much higher than the true value, or the addition of spurious $CH_4$ or $CO_2$ in the atmosphere. This finding is in line with the results of J. Gomez Barrientos et al. (2023) and F. Wang et al. (2022), which also show the difficulties of fitting the effects of the red edge with a flat albedo model and how this can cause biases in the posteriors of various gas abundances. This paper takes this one step further by showing how the biases fluctuate with the amount of vegetation in view. We therefore recommend the use of wavelength-dependent surface albedo, like the formulation presented here, to properly characterize Earth-like planets using reflection spectroscopy.

### 4.3. Biases Due to Ice Clouds

The test run retrievals in Appendix A indicate the possibility of strong biases arising from not accounting for the presence of ice cloud particles in the retrieval model. The input spectra shown here all include both liquid water and water ice particles in the clouds, whereas the retrieval model only includes liquid water particles. Figure 10 shows that when including the ice cloud particles in the input spectrum the retrieval returns an atmosphere with almost 20% $CO_2$. This is due to the different slope in reflectance that the ice and liquid particles have between 1.4 and 1.7 $\mu$m, in the same region where $CO_2$ features are present. However, despite also including ice particles in the input spectra, none of the other retrievals analyzed for this study show the same behavior, so the impact of the cloud types and their specific albedo spectra on the retrieval of realistic planet spectra may not be as severe as shown in the idealized test case. Despite this, it is still worth noting that this could be a potential source of biases, and it is recommended as an avenue for future study.

### 4.4. “Snapshots” versus Real Observations

The LUVOIR and HabEx mission concept studies suggest that for some Earth-like planets, exposure times of several days may be needed to achieve the desired SNR (The LUVOIR Team 2019; B. S. Gaudi et al. 2020). This means that we will likely not have the brief snapshots of different parts of the surface shown in this paper. However, as shown in Section 3.4, retrievals on the time-averaged spectra show a lot of the same surface-dependent behavior and are very similar to several of the “snapshot” spectra. With differing proportions of the various surface types, these snapshot spectra serve as good representative test cases for potential exo-Earths, which may have similar distributions when averaged over the entire surface. At least for an Earth twin, averaging over a period longer than the rotation period would not completely smear out the spectral features from heterogeneous and wavelength-dependent surface albedo. This constitutes, to our knowledge, the first proof of concept of detecting the red edge by spectroscopy from a complex terrestrial planet.

## 5. Conclusion

Previous research has proven that reflectance spectroscopy can be a powerful tool for the characterization of Earth-like exoplanets (Y. K. Feng et al. 2018; M. Damiano & R. Hu 2022). Thus far, the majority of studies have only considered 1D, spectrally flat surface albedo models. The effect of the wavelength-dependent, heterogeneous surface of an Earth-like planet on reflectance spectroscopy has been relatively understudied. This study aims to help close that gap by determining how the heterogeneous surface impacts retrievals of reflectance spectra.

To investigate this, we utilized PSG to generate full 3D, disk-integrated spectra of the Earth at various times of day (i.e., with differing surface features and cloud coverage; V. Kofman et al. 2024). We then performed retrievals on these spectra with ExoReL$^{\Re}$—extending the model by including the ability to fit the wavelength-dependent albedo with a simple step function—with the goal of determining how well an exo-Earth could be characterized when considering realistic surfaces.

The results of this study show the ability to detect signatures of the vegetation red edge from reflectance spectra when considering a complex, heterogeneous planetary surface (provided there is enough land/vegetation in view). We successfully retrieve the red edge even when considering integration times longer than the planet's rotation period. This lends support to the idea of utilizing the red edge as a biosignature in the context of the Habitable Worlds Observatory and opens up the potential of detecting other surface biosignatures.

We also find that flat albedo models can often return very biased posteriors on spectra from full 3D models. The retrieval results are quite sensitive to the surface features in view during the observation. When using a flat albedo model, there is a risk that the retrieval may try to fit the wavelength-dependent surface features using the clouds or gas abundances. With more land in view, the retrievals tend to underestimate the abundance of $O_2$ and $H_2O$, misidentify clouds, or infer spurious $CH_4$ or $CO_2$. Including a step function for the wavelength-dependent albedo mitigates many of these biases, producing much more accurate retrieval results. These results suggest that we need to seriously consider the effects of wavelength-dependent surface albedos to properly characterize Earth-like planets.

This study further demonstrates both the capabilities and limitations of retrievals for the characterization of rocky planets imaged in reflected light. It addresses key scientific questions related to the effects of complex, heterogeneous surface features and cloud type and distribution on spectral data and aims to provide valuable insights for future exoplanet imaging missions. The results will contribute to the refinement of retrieval techniques and support the goals of the Astro2020 decadal survey, advancing our ability to detect and characterize Earth-like exoplanets.

## Acknowledgments

We thank Vanessa Bailey for contributions to the noise model used in this study and we thank Luis Welbanks for the useful discussion on the LOO-CV analysis. The High Performance Computing resources used in this investigation were provided by funding from the JPL Information and Technology Solutions Directorate. This work was supported in part by the NASA grant #80NM0018F0612 in tandem with the Scialog 2023 "Signatures of Life in the Universe." This research was carried out at the Jet Propulsion Laboratory, California Institute of Technology, under a contract with the National Aeronautics and Space Administration (80NM0018D0004). V.K., G.L.V., and PSG are supported by the Goddard's Sellers Exoplanet Environments Collaboration (SEEC) and the Exoplanets Spectroscopy Technologies (ExoSpec) programs, which are part of NASA Science Division's Internal Scientist Funding Model.

*Software:* ExoReL$^{\Re}$ (GitHub, M. Damiano & R. Hu 2022), PSG (G. L. Villanueva et al. 2018, 2022).

## Appendix A
## Test Run Results

The in-depth results from the validation retrievals (outlined in Section 2.3) are discussed here.

The resulting posteriors for the clear and cloudy cases are shown in Figures 9 and 10, respectively. For the clear case, the posteriors are very close to the truth for all the parameters. $H_2O$, $O_3$, and $N_2$ are all retrieved to within $1\sigma$. There is no constraint on $CO_2$, which is to be expected as there are no strong $CO_2$ absorption features in this range. The retrieval results are also consistent with there being no $CH_4$. The retrieval identifies that there are no clouds present (returning a cloud fraction of 2% and no fit on the other cloud parameters), and is even able to correctly determine the radius and the surface albedo, which are known to be difficult to fit together (Y. K. Feng et al. 2018). The $O_2$ is slightly underestimated, but it is of the correct order of magnitude, and $O_2$ is very difficult to constrain at this low spectral resolution as there is only one data point in the $O_2$ absorption feature at 0.76 $\mu$m.

For the cloudy case, the retrieval returns an $O_2$-dominated atmosphere with a large percentage of $CO_2$. The overestimation of $O_2$ is likely because the clouds obscure part of the Rayleigh slope, which—together with the $O_2$ absorption—is an important feature used to distinguish between $N_2$ and $O_2$. Without the information on $N_2$, the retrieval defaults to $O_2$ being the dominant gas since that would make it easier to fit the spectral feature of $O_2$. This result is consistent with previous studies, which have shown that it can be difficult to determine the dominant gas of the atmosphere with reflection spectroscopy, and in particular to differentiate between $N_2$ and $O_2$ in the case of modern Earth-like planets (S. Hall et al. 2023; E. Alei et al. 2024; M. Damiano et al. 2025). The $CO_2$ abundance is also significantly overestimated and has a very narrow posterior, showing a very confident detection. $CH_4$ shows a slight peak at $10^{-4}$, but the results are also consistent with no $CH_4$. Several of the other parameters are also biased to higher or lower values. The fit on the clouds is not exact, but fairly good.

The issues with this retrieval are due to differences in the cloud models between PSG and ExoReL$^{\Re}$. PSG includes ice cloud particles, but ExoReL$^{\Re}$ does not. The reflectance of liquid water clouds has a sharp increase in extinction blueward of 1.5 $\mu$m, whereas ice clouds have a more gradual increase between 1.4 and 1.7 $\mu$m, which causes the appearance of a drop in the contrast in this range. In Figure 10, it can be seen that the best-fit model does not fit the data well in this region for this reason. This is also the same spectral range where $CH_4$ and $CO_2$ have absorption features, so the retrieval fits this signature by increasing the VMR of these molecules. Ice clouds have a higher reflectance overall, so to compensate for this the retrieval returns a radius that is slightly higher than the truth.

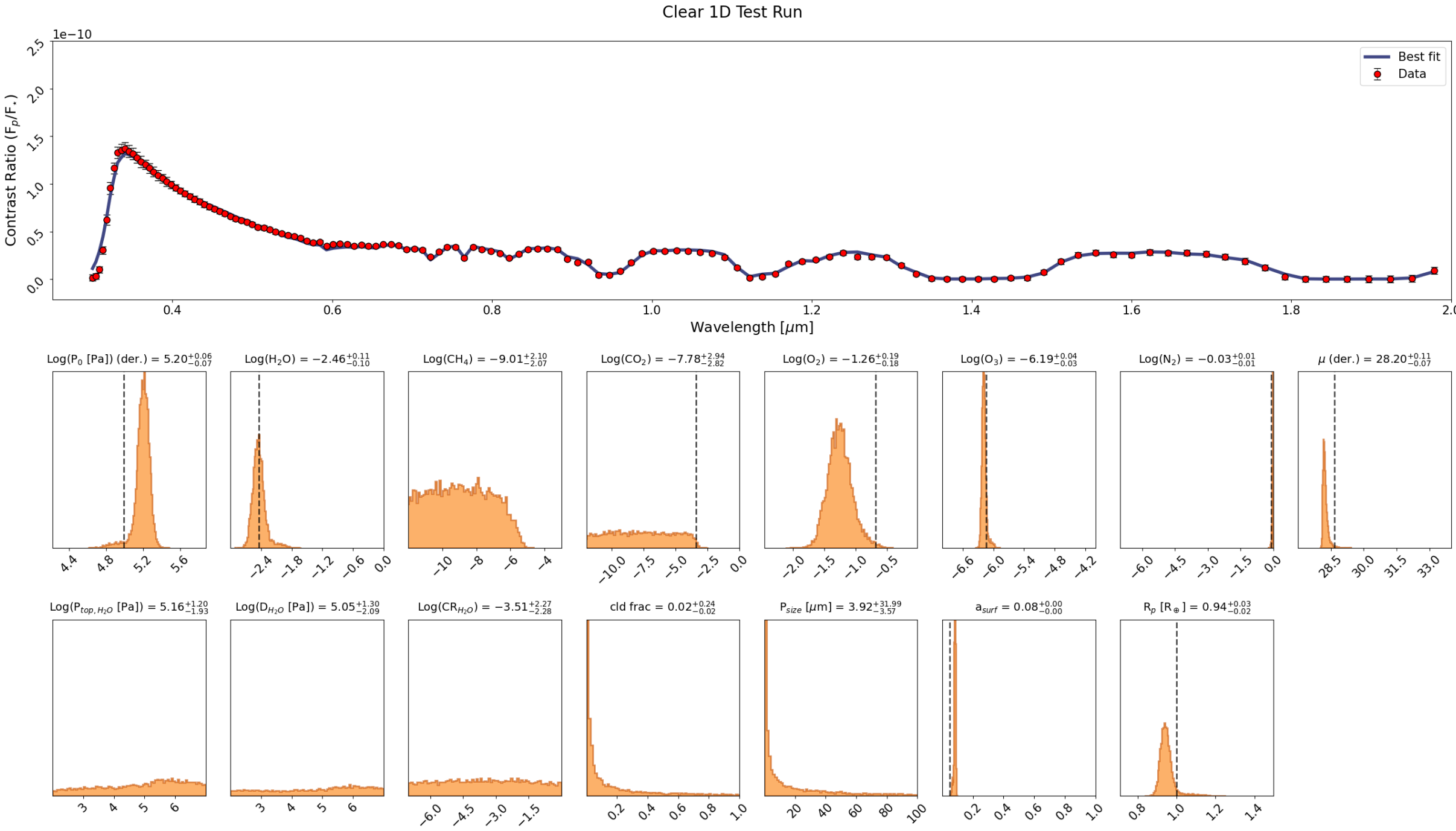

**Figure 9.** Best-fit spectrum and posteriors for a retrieval on a clear sky spectrum generated by PSG with a spectrally flat surface albedo. The retrieved parameters are almost all within $3\sigma$ of the true values, except for $O_2$ (at $4\sigma$ away), which is very difficult to fit at this low spectral resolution. There is only an upper bound for the $CO_2$, which is to be expected as there are no strong $CO_2$ features in this spectrum. The retrieval also correctly identifies that there are no clouds. These results demonstrate that ExoReL$^{\Re}$ can successfully retrieve on PSG spectra absent clouds and 3D effects.

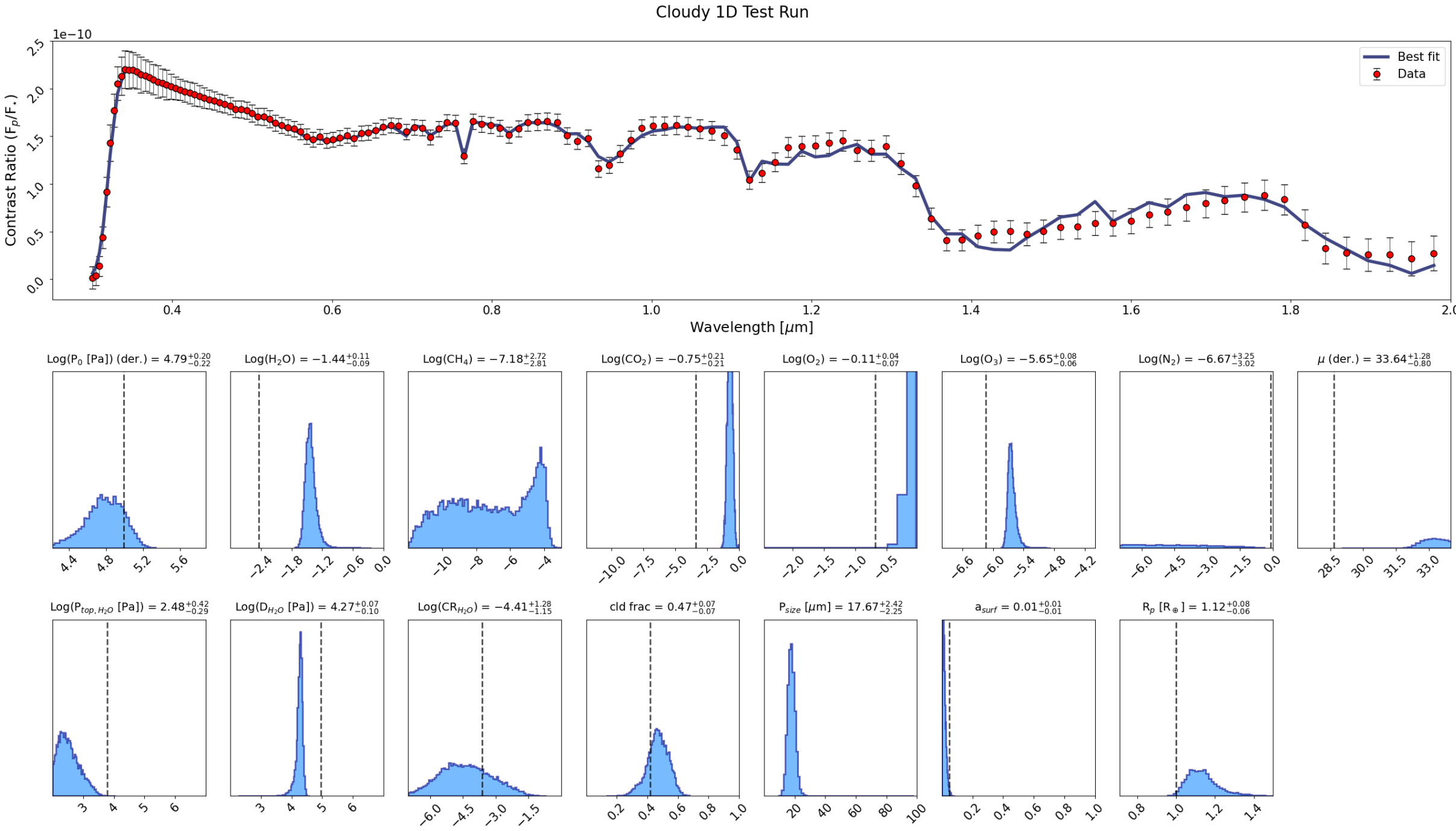

**Figure 10.** Best-fit spectrum and posteriors for a retrieval on a cloudy spectrum generated by PSG with a spectrally flat surface albedo. Due to differences in the cloud descriptions between the forward models, the retrieval returns heavily biased results. The retrieval incorrectly returns an $O_2$ atmosphere with 20% $CO_2$, which shows a very confident detection despite $CO_2$ only having very weak spectral features that are not easily detectable in this wavelength range. The best-fit spectrum also fails to fit the data between 1.4 and 1.7 $\mu$m, due to ice cloud particles (included in PSG but not ExoReL$^{\Re}$), which have a slightly different slope in this region.

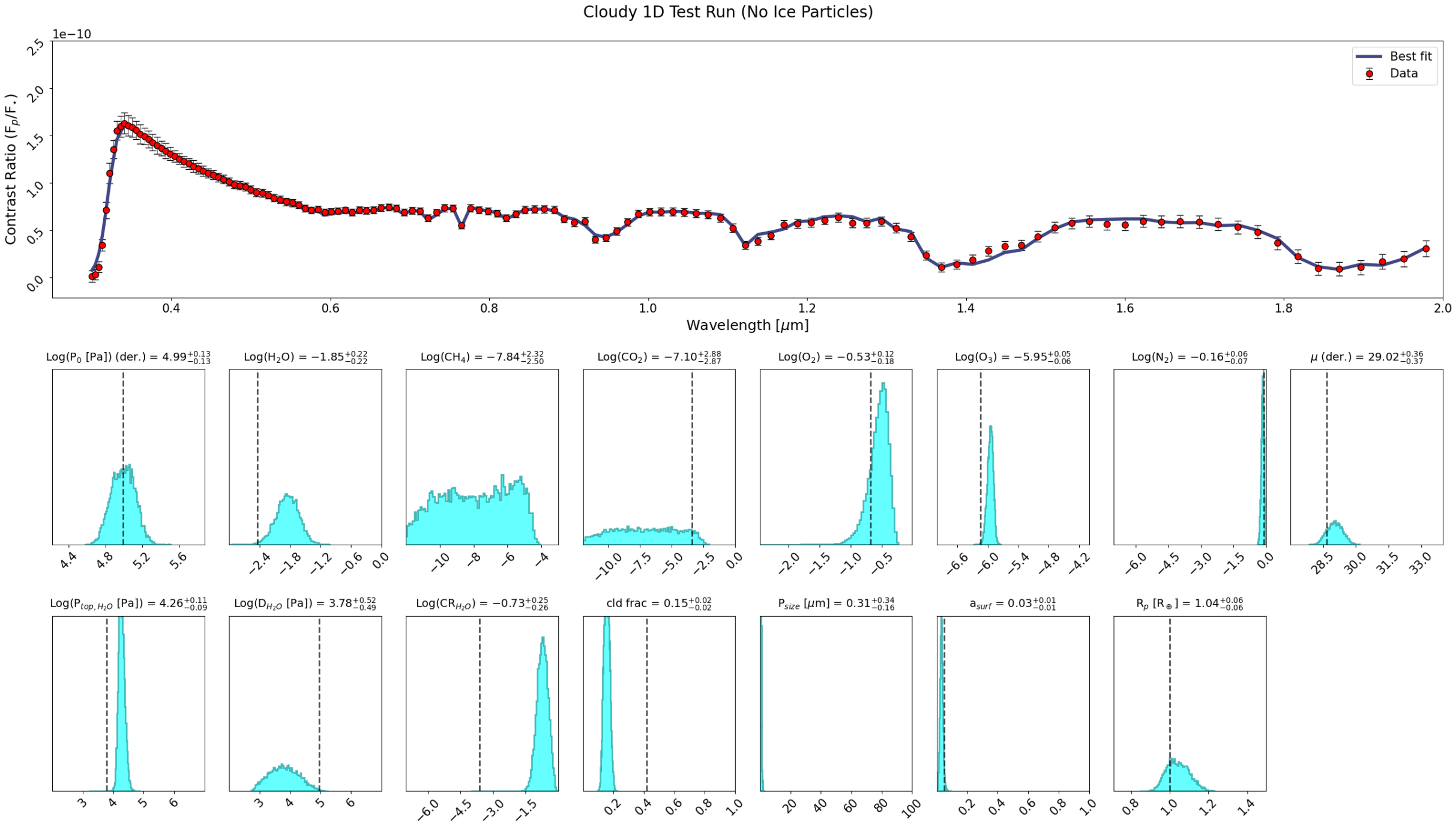

**Figure 11.** Same as Figure 10, except the ice cloud particles have been removed from the PSG model. Comparing to the previous results the retrieval performs much better here. The best-fit spectrum now fits the data correctly, the spurious detections of $CH_4$ and $CO_2$ have disappeared, and the other gas abundances are now all fit accurately.

To account for these effects, an additional test run retrieval was performed where ice particles were no longer included in the PSG simulation. The results of this are shown in Figure 11. Without the ice cloud particles, the best-fit spectrum now fits much better to the input. Additionally, the posteriors all get much closer to the truth. The gas abundances are now all within $3\sigma$ (except $O_3$ at $4\sigma$, which is likely due to the fact the $O_3$ varies vertically in the PSG model but is constant in ExoReL$^{\Re}$). The spurious detection of $CO_2$ has also disappeared, and now only has an upper bound. The radius and surface albedo are fit correctly. The cloud parameters and cloud fraction still vary from the truth. Some residual differences could still exist in the handling of clouds since the models utilize different multiple-scattering solvers, with PSG employing PSGDORT (G. L. Villanueva et al. 2022) based on DISORT (K. Stamnes et al. 1988), and ExoReL$^{\Re}$ using a two-stream source function (R. Hu 2019).

The fact that the clear case is very accurate shows that ExoReL$^{\Re}$ is capable of accurately retrieving planetary and atmospheric parameters from spectra generated by PSG. The differences in the cloud models introduce some additional biases, which makes determining the effect of solely the surface more difficult. However, by comparing between cloudy retrievals with different types of surfaces visible and seeing where the biases differ it is possible to assess the effect of the surface itself in tandem with the clouds.

## Appendix B
## Partial Pressure versus Center Log Ratio

Center log ratio (CLR) has historically been a common way to fit gas abundances during retrievals of small planets (B. Benneke & S. Seager 2012; M. Damiano & R. Hu 2021), and is the approach previously used for studies with ExoReL$^{\Re}$ (M. Damiano & R. Hu 2022; M. Damiano et al. 2023). In this work, we instead choose to fit the gases in the partial pressure space, with

$$P_0 = \sum_n \delta P_{\text{gas}_n} \tag{B1}$$

$$\delta P_{\text{gas}_n} = P_0 \times \text{VMR}_{\text{gas}_n} \tag{B2}$$

where $\delta P_{\text{gas}_n}$ is the partial pressure of the $n$th gas, $\text{VMR}_{\text{gas}_n}$ is the volume mixing ratio of the $n$th gas, and $P_0$ is the surface pressure (M. Damiano et al. 2025).

This offers many advantages compared to the CLR space: there is no need for a filling gas, as all gases can be fit independently; the surface pressure can be retrieved together with the gases, rather than as a separate parameter; and the transformation from partial pressure to VMR is much easier than for CLR.

Because CLR and partial pressure are transformations of each other, the results when fitting with either should be equivalent. To confirm that the results are identical, we performed an additional retrieval with the gases fit by CLR to serve as a validation. The spectrum used is the clear 1D test run spectrum generated with PSG as described in Section 2.3, and the retrieval setup is as described in Section 2.2, except with $N_2$ as a filling gas, and the gas priors are as described by M. Damiano & R. Hu (2021). This is a relatively simple case, which serves as a good validation exercise: without other confounding factors such as clouds the retrieval is more sensitive to the gas abundances, so any differences between the partial pressure and CLR methods will be more obvious.

The full results for the partial pressure fitting are shown in Figure 9. The comparison between the posteriors for the partial pressure and CLR is shown in Figure 12. The posteriors for the surface pressure and the gas abundances are all identical, showing that these two methods are equivalent.

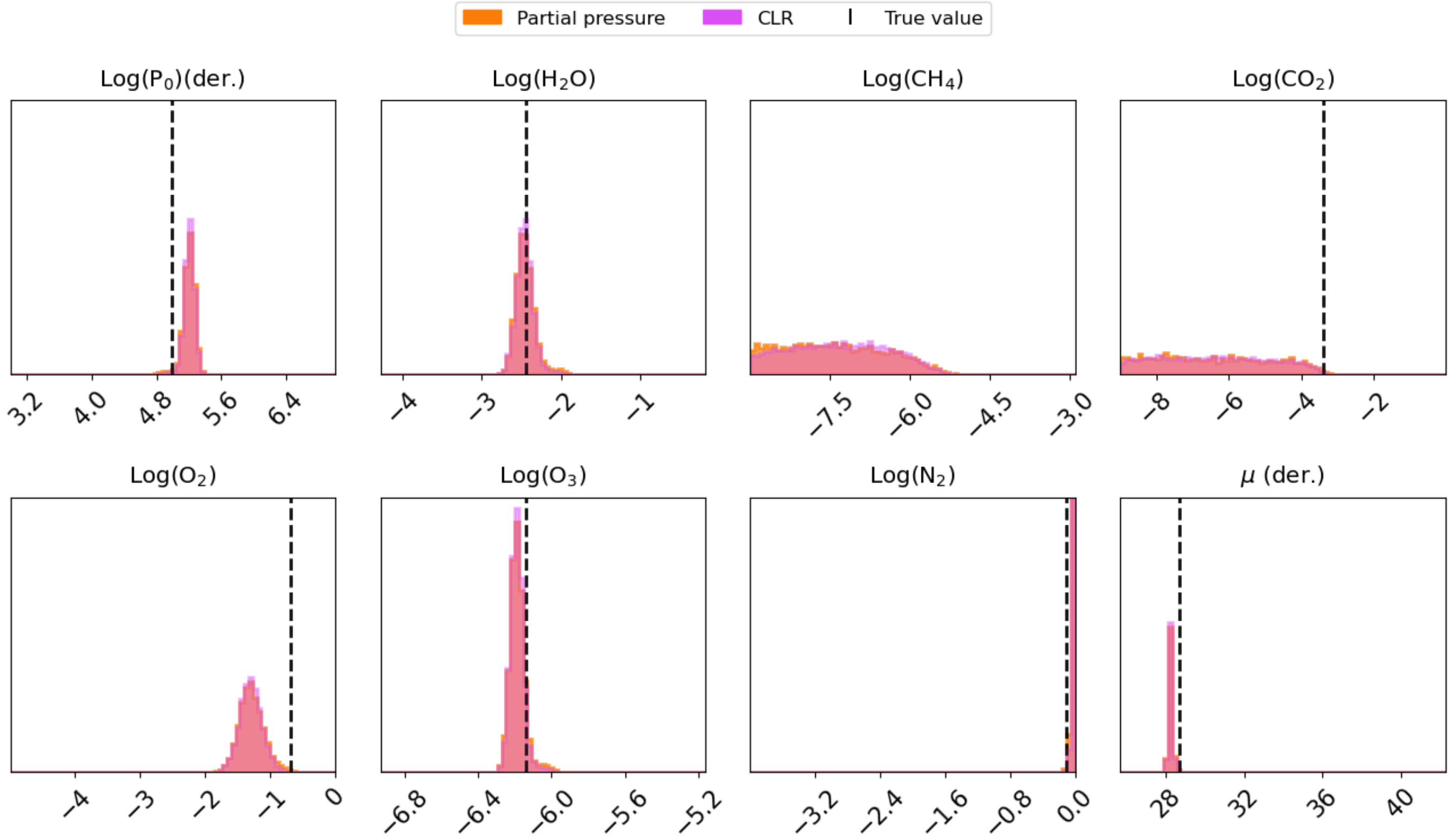

**Figure 12.** Comparison between fitting the gas abundances in the partial pressure space and the CLR space. This test is performed on the clear 1D case presented in Section 2.3. The posteriors for all the parameters are identical, demonstrating that the two methods are equivalent.

## ORCID iDs

Zachary Burr https://orcid.org/0009-0000-4845-3613
Mario Damiano https://orcid.org/0000-0002-1830-8260
Vincent Kofman https://orcid.org/0000-0002-5060-1993
Renyu Hu https://orcid.org/0000-0003-2215-8485
Geronimo L. Villanueva https://orcid.org/0000-0002-2662-5776